\documentstyle[12pt]{article}\newlength\Cscr\newlength\Csave\newlength
\Ctenthex\newlength\CFxsize\newlength
\CFxsizeps\newlength\CFsizemakebox\newlength\CFleftcrop\newlength
\CFrightcrop\newlength\CZtbldist\newlength\CZfigdist\newlength\CGDnum
\newlength\CGDtext\newcounter{Cscr}\newcounter
{CBcit}\newcounter{CElett}\newcounter{CFlette%
r}\newcounter{Ceqin%
dent}\newcounter{CBtnc}\newcounter{CBtntc}%
\newcounter{CEht}%
\newcounter{CbsA}\newcounter{CbsB}\newcounter{CbsC}\newcounter{C%
bsD}
\hoffset\Cscr
\voffset\Cscr
\protect
\begin{document}\setcounter{figure}{1}\setcounter{table}%
{0}\renewcommand\theequation{\arabic{equation}}\renewcommand
\thetable{\arabic{table}}\renewcommand\thefigure{\arabic{figure}%
}\renewcommand\thesection{\arabic{section}}\renewcommand
\thesubsection{\arabic{section}.\arabic{subsection}}\renewcommand
\thesubsubsection{\arabic{section}.\arabic{subsection}.\arabic{s%
ubsubsection}}\setcounter{CEht}{10}\setcounter{CbsA}{1}%
\setcounter{CbsB}{1}\setcounter{CbsC}{1}\setcounter{CbsD}{1}{%
\centering{\protect\mbox{}}\\*[\baselineskip]{\large\bf A strang%
e property of lattices with an even number of sites}\\*}{%
\centering{\protect\mbox{}}\\John P.~Costella\vspace{1ex}\\*}{%
\centering{\small\mbox{}\protect\/{\protect\em Mentone Grammar, 
63 Venice Street, Mentone, Victoria 3194, Australia\protect\/}}%
\\}{\centering{\small\mbox{}\protect\/{\protect\em jpcostella@ho%
tmail.com;\hspace{1ex} jpc@mentonegs.vic.edu.au;\hspace{1ex} www%
.ph.unimelb.edu.au/$\sim$jpc\protect\/}}\\}{\centering{\protect
\mbox{}}\\(8\ April 2004\vspace{1ex})\\}\par\vspace\baselineskip
\setlength{\Csave}{\parskip}{\centering\small\bf Abstract\\}%
\setlength{\parskip}{-\baselineskip}\begin{quote}\setlength{%
\parskip}{0.5\baselineskip}\small\noindent By examining the beha%
viour\ of the ``SLAC'' lattice derivative operators, it is found 
that lattices with an even number of sites have a somewhat stran%
ge self-consistency requirement for extra structure in the spati%
al derivative operator, which is not needed by lattices having a%
n odd number of sites, and which is not at all obvious from a fi%
rst-principles derivation. The general implications of this extr%
a required structure are not, as yet, completely clear. \end{quo%
te}\setlength{\parskip}{\Csave}\par\refstepcounter{section}%
\vspace{1.5\baselineskip}\par{\centering\bf\thesection. Introduc%
tion\\*[0.5\baselineskip]}\protect\indent\label{sect:Introductio%
n}In a previous pair of papers~[\ref{cit:Costella2002a},\,\ref{c%
it:Costella2002b}]{} I provided a pedagogically instructive deri%
vation of the ``SLAC'' spatial derivative operators of \protect
\ref{au:Drell1976a}~[\ref{cit:Drell1976a},\,\ref{cit:Drell1976b}%
].{} Their approach was to ensure that the Fourier transform of 
the spatial derivative operator takes on its ideal form in momen%
tum space, within the first Brillouin zone---namely, that the Fo%
urier transform of $(-i\partial/\partial x)_{\mbox{\scriptsize S%
LAC}}$ be identically equal to $p$. This is in contrast to what 
we obtain with $-i$ times the na\-\"\i ve, nearest-neighbour\ la%
ttice derivative operator, which has Fourier transform of $\sin(%
pa)/a$ (which is only approximately linear near \raisebox{0ex}[1%
ex][0ex]{$\protect\displaystyle p=0$}, and possesses the infamou%
s spurious zero at the Brillouin zone boundary \raisebox{0ex}[1e%
x][0ex]{$\protect\displaystyle p=\pm\pi/a$} that fundamentally c%
auses the fermion doubling problem).\par My goal in those earlie%
r papers was to present a new proposal for a computationally eff%
icient implementation of the SLAC derivative operators, by emplo%
ying a stochastic philosophy. However, the pedagogical derivatio%
n of the SLAC operators, in closed form, prompted me to first ch%
eck their expected fundamental properties in some practical test 
calculations.\par Of particular interest to me were the expressi%
ons for the SLAC derivative operators for lattices with an even 
number of sites, because \protect\ref{au:Drell1976a} restricted 
their attention to lattices with an odd number of sites only. (T%
hey furthermore did not obtain an expression for the \mbox{}%
\protect\/{\protect\em first\protect\/}-derivative operator in c%
losed form, in position space, even for lattices with an odd num%
ber of sites; rather, they simply represented it~[\ref{cit:Drell%
1976b}] as the inverse Fourier transform of~$p$.) My intention w%
as to confirm the properties of those closed-form expressions ob%
tained in [\ref{cit:Costella2002a},\,\ref{cit:Costella2002b}]\ t%
hat were not obtained by \protect\ref{au:Drell1976a}, to ensure 
that no mathematical or algebraic slips had been made in their d%
erivation.\par These confirmations were simple to obtain for the 
expressions for lattices with an odd number of sites. However, o%
nce the number of sites was changed to an even number, the confi%
rmation cross-checks continually and inexplicably failed.\par Al%
though it has taken a somewhat embarrassingly long time for me t%
o find the time to dissect the results and discover ``where the 
wheels fell off''---and despite the fact that the source of the 
``problem'' is blindingly obvious in retrospect---I believe that 
it is of sufficient intrinsic interest (and, quite possibly, fun%
damental importance) to those employing lattices with an even nu%
mber of sites that a detailed discussion of the issue is warrant%
ed. This is the focus of the current paper.\par\refstepcounter{s%
ection}\vspace{1.5\baselineskip}\par{\centering\bf\thesection. T%
he general philosophy of the previous two papers\\*[0.5%
\baselineskip]}\protect\indent\label{sect:Philosophy}In [\ref{ci%
t:Costella2002a}] I provided a general philosophical framework f%
or what might be argued to be an ``optimal'' construction of the 
description of a physical system on a lattice. The basic idea is 
to start with the physical description that would be appropriate 
if both position space and momentum space are continua of infini%
te extent. This description then needs to be limited with a ``lo%
w pass'' filter in momentum space, which cuts off anything outsi%
de (what will be) the first Brillouin zone. This prepares the ph%
ysical system for ``sampling'' onto the lattice in position spac%
e. However, \mbox{}\protect\/{\protect\em before we actually per%
form this sampling\protect\/}, we obtain expressions for all the 
operators that we shall require (in our case of interest here th%
e spatial derivative operators). Only \mbox{}\protect\/{\protect
\em then\protect\/} do we sample everything onto the lattice in 
position space.\par At this stage, the physical system has been 
sampled onto an infinite lattice in position space, which implie%
s an essential momentum-space description that is still a contin%
uum, but contained within the first Brillouin zone. The second p%
art of the process is to consider the more practically useful ca%
se when the position-space lattice is limited in extent, with pe%
riodic boundary conditions applied. This effectively ``samples'' 
the description in momentum space, as it is converted into a seq%
uence of Dirac delta functions, which can in turn be summarised 
by the set of their co\-efficients, which effectively describes 
a lattice in momentum space. A one-dimensional lattice of $N$ si%
tes in position space then corresponds to a one-dimensional latt%
ice of $N$ sites in momentum space; the two descriptions are rel%
ated by the discrete Fourier transform.\par The difficult part o%
f this second stage is to apply the periodic boundary conditions 
to the operators found in the first stage; effectively, the infi%
nite-lattice operators have to be ``wrapped around'' the finite 
lattice an infinite number of times, with the contributions of a%
ll the ``layers'' to any given finite-lattice site summed up. Th%
is second stage was described in [\ref{cit:Costella2002b}] for t%
he case of the spatial derivative operators, where it was shown 
that the infinite sums could be performed in closed form, and in%
deed yield rather simple expressions.\par The great advantage of 
this general constructional framework is that operators defined 
in this way in position space will automatically possess a corre%
ct (continuum) representation in momentum space---that is, at le%
ast within the first Brillouin zone and at the momentum values c%
orresponding to the momentum-space lattice sites. In this sense 
they are as ``optimal'' as one can possibly achieve with a latti%
ce of the given size, in that they represent their continuum cou%
nterparts with the maximum ``fidelity'' possible, without unnece%
ssary distortion or artefacts. Clearly, in the case of the spati%
al derivative operators, the process must yield the ``SLAC'' ope%
rators of \protect\ref{au:Drell1976a}, because their prescriptio%
n was to simply \mbox{}\protect\/{\protect\em start\protect\/} w%
ith this momentum-space fidelity, and determine the correspondin%
g position-space operators (to the extent that they \mbox{}%
\protect\/{\protect\em did\protect\/} actually determine these o%
perators) by means of the inverse discrete Fourier transform.\par
\refstepcounter{section}\vspace{1.5\baselineskip}\par{\centering
\bf\thesection. Expected properties of the SLAC derivative opera%
tors\\*[0.5\baselineskip]}\protect\indent\label{sect:ExpectedSLA%
C}The momentum-space fidelity of the SLAC derivative operators (%
and by the description ``SLAC'' I shall henceforth refer to all 
of the spatial derivative operators obtained in [\ref{cit:Costel%
la2002b}] by means of what is effectively the ``SLAC'' philosoph%
y, regardless of whether or not \protect\ref{au:Drell1976a} actu%
ally obtained closed-form expressions for them in position space%
) means that there are a number of properties of the continuum o%
perators that we would expect to be obeyed by their SLAC counter%
parts on the lattice.\par Chief amongst these is the property th%
at the second-derivative operator should be simply equivalent to 
two successive applications of the first-derivative operator. Th%
is can be seen most clearly in momentum space. The application o%
f the spatial derivative operator to a function in position spac%
e is equivalent to the convolution of the said function with the 
``position space representation'' of the spatial derivative oper%
ator (which can be obtained by applying the spatial derivative o%
perator to the negative of the lattice representation of the Dir%
ac delta function at the origin). By the principles of Fourier t%
heory, a convolution in position space is equivalent to a simple 
multiplication in momentum space. Now, since the momentum-space 
representation of the SLAC first-derivative operator is simply $%
-ip$, applying the first-derivative operator twice is equivalent 
to multiplying by \raisebox{0ex}[1ex][0ex]{$\protect\displaystyle
(-ip)^2\equiv-{p}{}^{\raisebox{-0.25ex}{$\scriptstyle\:\!2$}}$} 
in momentum space. But this is precisely the momentum-space repr%
esentation of the SLAC \mbox{}\protect\/{\protect\em second%
\protect\/}-derivative operator (by definition). Thus, a sequenc%
e of two successive applications of the SLAC first-derivative op%
erator is equivalent to the application of the SLAC second-deriv%
ative operator, as is of course true in the continuum descriptio%
n.\par This property may seem to be almost trivial, until one re%
alises that it is \mbox{}\protect\/{\protect\em not\protect\/} o%
beyed by the na\-\"\i ve\ first- and second-derivative operators%
. This is seen most quickly by remembering that the na\-\"\i ve\ 
first-derivative operator possesses a spurious zero at the Brill%
ouin zone boundary (that is responsible for fermion doubling in 
lattice gauge theory), whereas the na\-\"\i ve\ second-derivativ%
e operator possesses no such zero~[\ref{cit:Costella2002a}]. Alt%
ernatively (or, really, equivalently), the period of the momentu%
m-space representation of the na\-\"\i ve\ first-derivative oper%
ator is only half that of the na\-\"\i ve\ second-derivative ope%
rator. All that these na\-\"\i ve\ operators guarantee is that 
\mbox{}\protect\/{\protect\em as\protect\/} \raisebox{0ex}[1ex][%
0ex]{$\protect\displaystyle p\rightarrow0$} the first-derivative 
operator approaches $ip$ and the second-derivative operator appr%
oaches \raisebox{0ex}[1ex][0ex]{$\protect\displaystyle-{p}{}^{%
\raisebox{-0.25ex}{$\scriptstyle\:\!2$}}$}, so that the latter i%
s only approximately the square of the former, as \raisebox{0ex}%
[1ex][0ex]{$\protect\displaystyle p\rightarrow0$}; and since thi%
s relationship does not extend over all $p$, it means that there 
is no exact relationship in position space between two successiv%
e applications of the na\-\"\i ve\ first-derivative operator, an%
d a single application of the na\-\"\i ve\ second-derivative ope%
rator.\par Now, it is one thing to determine, theoretically, tha%
t two successive applications of the SLAC first-derivative opera%
tor should be equivalent to a single application of the SLAC sec%
ond-derivative operator. It is quite another to verify this prop%
erty, computationally, for actual lattices of various sizes. It 
is in this task that my own computations both succeeded and fail%
ed, as will now be described; and it is the failures that provid%
e the phenomenon that is of fundamental interest to us.\par
\refstepcounter{section}\vspace{1.5\baselineskip}\par{\centering
\bf\thesection. Summary of the SLAC derivative operator expressi%
ons\\*[0.5\baselineskip]}\protect\indent\label{sect:SLAC}The $m$%
-th order one-dimensional SLAC spatial derivative operator, ${%
\protect\it\Delta\!\:}_{\mbox{\scriptsize SLAC}}^{(m)}$, is defi%
ned by its action on an arbitrary function \raisebox{0ex}[1ex][0%
ex]{$\protect\displaystyle f_{\:\!\!n}$} defined on the sites 
\raisebox{0ex}[1ex][0ex]{$\protect\displaystyle x_n\equiv na$} o%
f a position-space lattice containing $N$ sites: \setcounter{Ceq%
indent}{0}\protect\begin{eqnarray}\protect\left.\protect\begin{a%
rray}{rcl}\protect\displaystyle\hspace{-1.3ex}&\protect
\displaystyle({\protect\it\Delta\!\:}_{\mbox{\scriptsize SLAC}}^%
{(m)}\,f)_n\equiv\sum_{n'}\,{}^{(m)\!}c_{n'\!-n}^N\,f_{\:\!\!n'}%
,\setlength{\Cscr}{\value{CEht}\Ctenthex}\addtolength{\Cscr}{-1.%
0ex}\protect\raisebox{0ex}[\value{CEht}\Ctenthex][\Cscr]{}%
\protect\end{array}\protect\right.\protect\label{eq:SLAC-Coeffic%
ients}\protect\end{eqnarray}\setcounter{CEht}{10}where the co\-e%
fficients \raisebox{0ex}[1ex][0ex]{$\protect\displaystyle{}^{(m)%
\!}c_n^N$} were obtained in closed form in [\ref{cit:Costella200%
2a}]: \setcounter{CElett}{0}\protect\refstepcounter{equation}%
\protect\label{eq:SLAC-SLAC}\renewcommand\theequation{\arabic{eq%
uation}\alph{CElett}}\setcounter{Ceqindent}{0}\protect\begin{eqn%
array}\setcounter{CEht}{30}{}^{(1)\!}c_0^N\hspace{-1.3ex}&%
\displaystyle=&\hspace{-1.3ex}0,\protect\stepcounter{CElett}%
\addtocounter{equation}{-1}\protect\label{eq:SLAC-SLAC-1-0}%
\setlength{\Cscr}{\value{CEht}\Ctenthex}\addtolength{\Cscr}{-1.0%
ex}\protect\raisebox{0ex}[\value{CEht}\Ctenthex][\Cscr]{}\\*[0ex%
]\protect\displaystyle{}^{(1)\!}c_{n\neq0}^{\mbox{\scriptsize$N%
\!$ odd}}\hspace{-1.3ex}&\displaystyle=&\hspace{-1.3ex}\mbox{$%
\protect\displaystyle\protect\frac{(-1)^{n+1\!}}{na}$}\,\alpha
\csc\alpha,\protect\stepcounter{CElett}\addtocounter{equation}{-%
1}\protect\label{eq:SLAC-SLAC-1-n-odd}\setlength{\Cscr}{\value{C%
Eht}\Ctenthex}\addtolength{\Cscr}{-1.0ex}\protect\raisebox{0ex}[%
\value{CEht}\Ctenthex][\Cscr]{}\\*[0ex]\protect\displaystyle{}^{%
(1)\!}c_{n\neq0}^{\mbox{\scriptsize$N\!$ even}}\hspace{-1.3ex}&%
\displaystyle=&\hspace{-1.3ex}\mbox{$\protect\displaystyle
\protect\frac{(-1)^{n+1\!}}{na}$}\,\alpha\cot\alpha,\protect
\stepcounter{CElett}\addtocounter{equation}{-1}\protect\label{eq%
:SLAC-SLAC-1-n-even}\setlength{\Cscr}{\value{CEht}\Ctenthex}%
\addtolength{\Cscr}{-1.0ex}\protect\raisebox{0ex}[\value{CEht}%
\Ctenthex][\Cscr]{}\\*[0ex]\protect\displaystyle{}^{(2)\!}c_0^{%
\mbox{\scriptsize$N\!$ odd}}\hspace{-1.3ex}&\displaystyle=&%
\hspace{-1.3ex}-\mbox{$\protect\displaystyle\protect\frac{{\pi}{%
}^{\raisebox{-0.25ex}{$\scriptstyle2$}}}{3a^2}$}\!^{\!}\protect
\left({}^{\:\!}1-\mbox{$\protect\displaystyle\protect\frac{1}{N^%
2}$}\!\protect\right)\!^{\!},\protect\stepcounter{CElett}%
\addtocounter{equation}{-1}\protect\label{eq:SLAC-SLAC-2-0-odd}%
\setlength{\Cscr}{\value{CEht}\Ctenthex}\addtolength{\Cscr}{-1.0%
ex}\protect\raisebox{0ex}[\value{CEht}\Ctenthex][\Cscr]{}\\*[0ex%
]\protect\displaystyle{}^{(2)\!}c_0^{\mbox{\scriptsize$N\!$ even%
}}\hspace{-1.3ex}&\displaystyle=&\hspace{-1.3ex}-\mbox{$\protect
\displaystyle\protect\frac{{\pi}{}^{\raisebox{-0.25ex}{$%
\scriptstyle2$}}}{3a^2}$}\!^{\!}\protect\left({}^{\:\!}1+\mbox{$%
\protect\displaystyle\protect\frac{2}{N^2}$}\!\protect\right)\!^%
{\!},\protect\stepcounter{CElett}\addtocounter{equation}{-1}%
\protect\label{eq:SLAC-SLAC-2-0-even}\setlength{\Cscr}{\value{CE%
ht}\Ctenthex}\addtolength{\Cscr}{-1.0ex}\protect\raisebox{0ex}[%
\value{CEht}\Ctenthex][\Cscr]{}\\*[0ex]\protect\displaystyle{}^{%
(2)\!}c_{n\neq0}^{\mbox{\scriptsize$N\!$ odd}}\hspace{-1.3ex}&%
\displaystyle=&\hspace{-1.3ex}\mbox{$\protect\displaystyle
\protect\frac{2(-1)^{n+1\!}}{n^2a^2}$}\,{\alpha}{}^{\raisebox{-0%
.25ex}{$\scriptstyle2\:\!\!$}}\csc\alpha\cot\alpha,\protect
\stepcounter{CElett}\addtocounter{equation}{-1}\protect\label{eq%
:SLAC-SLAC-2-n-odd}\setlength{\Cscr}{\value{CEht}\Ctenthex}%
\addtolength{\Cscr}{-1.0ex}\protect\raisebox{0ex}[\value{CEht}%
\Ctenthex][\Cscr]{}\\*[0ex]\protect\displaystyle{}^{(2)\!}c_{n%
\neq0}^{\mbox{\scriptsize$N\!$ even}}\hspace{-1.3ex}&%
\displaystyle=&\hspace{-1.3ex}\mbox{$\protect\displaystyle
\protect\frac{2(-1)^{n+1\!}}{n^2a^2}$}\,{\alpha}{}^{\raisebox{-0%
.25ex}{$\scriptstyle2\:\!$}}{\csc}{}^{\raisebox{-0.25ex}{$%
\scriptstyle2$}}\alpha,\protect\stepcounter{CElett}\addtocounter
{equation}{-1}\protect\label{eq:SLAC-SLAC-2-n-even}\setlength{%
\Cscr}{\value{CEht}\Ctenthex}\addtolength{\Cscr}{-1.0ex}\protect
\raisebox{0ex}[\value{CEht}\Ctenthex][\Cscr]{}\protect\end{eqnar%
ray}\setcounter{CEht}{10}\renewcommand\theequation{\arabic{equat%
ion}}where \setcounter{Ceqindent}{0}\protect\begin{eqnarray}%
\protect\left.\protect\begin{array}{rcl}\protect\displaystyle
\hspace{-1.3ex}&\protect\displaystyle\alpha\equiv\mbox{$\protect
\displaystyle\protect\frac{\pi n}{N}$}.\setlength{\Cscr}{\value{%
CEht}\Ctenthex}\addtolength{\Cscr}{-1.0ex}\protect\raisebox{0ex}%
[\value{CEht}\Ctenthex][\Cscr]{}\protect\end{array}\protect\right
.\protect\label{eq:SLAC-Alpha}\protect\end{eqnarray}\setcounter{%
CEht}{10}(The index \raisebox{0ex}[1ex][0ex]{$\protect
\displaystyle n'\!-n$} in (\protect\ref{eq:SLAC-Coefficients}) i%
s opposite to that used for a convolution; this is notationally 
convenient, because it means that the value $c_n$ is the co\-eff%
icient applied in the sum to the function value at the site whic%
h is at a distance $n$ from the site of differentiation.) The ex%
pressions (\protect\ref{eq:SLAC-SLAC-2-0-odd}) and (\protect\ref
{eq:SLAC-SLAC-2-n-odd}) were obtained by \protect\ref{au:Drell19%
76a} in [\ref{cit:Drell1976a}] (where the number of lattice site%
s was written \raisebox{0ex}[1ex][0ex]{$\protect\displaystyle2N+%
1$} rather than $N$). In the limit \raisebox{0ex}[1ex][0ex]{$%
\protect\displaystyle N\rightarrow0$} we obtain the infinite-lat%
tice expressions derived in [\ref{cit:Costella2002a}]: 
\setcounter{CElett}{0}\protect\refstepcounter{equation}\protect
\label{eq:SLAC-Infinite}\renewcommand\theequation{\arabic{equati%
on}\alph{CElett}}\setcounter{Ceqindent}{0}\protect\begin{eqnarra%
y}\setcounter{CEht}{30}{}^{(1)\!}c_0^\infty\hspace{-1.3ex}&%
\displaystyle=&\hspace{-1.3ex}0,\protect\stepcounter{CElett}%
\addtocounter{equation}{-1}\protect\label{eq:SLAC-Infinite-1-0}%
\setlength{\Cscr}{\value{CEht}\Ctenthex}\addtolength{\Cscr}{-1.0%
ex}\protect\raisebox{0ex}[\value{CEht}\Ctenthex][\Cscr]{}\\*[0ex%
]\protect\displaystyle{}^{(1)\!}c_{n\neq0}^\infty\hspace{-1.3ex}%
&\displaystyle=&\hspace{-1.3ex}\mbox{$\protect\displaystyle
\protect\frac{(-1)^{n+1\!}}{na}$},\protect\stepcounter{CElett}%
\addtocounter{equation}{-1}\protect\label{eq:SLAC-Infinite-1-n}%
\setlength{\Cscr}{\value{CEht}\Ctenthex}\addtolength{\Cscr}{-1.0%
ex}\protect\raisebox{0ex}[\value{CEht}\Ctenthex][\Cscr]{}\\*[0ex%
]\protect\displaystyle{}^{(2)\!}c_0^\infty\hspace{-1.3ex}&%
\displaystyle=&\hspace{-1.3ex}-\mbox{$\protect\displaystyle
\protect\frac{{\pi}{}^{\raisebox{-0.25ex}{$\scriptstyle2$}}}{3a^%
2}$},\protect\stepcounter{CElett}\addtocounter{equation}{-1}%
\protect\label{eq:SLAC-Infinite-2-0}\setlength{\Cscr}{\value{CEh%
t}\Ctenthex}\addtolength{\Cscr}{-1.0ex}\protect\raisebox{0ex}[%
\value{CEht}\Ctenthex][\Cscr]{}\\*[0ex]\protect\displaystyle{}^{%
(2)\!}c_{n\neq0}^\infty\hspace{-1.3ex}&\displaystyle=&\hspace{-1%
.3ex}\mbox{$\protect\displaystyle\protect\frac{2(-1)^{n+1\!}}{n^%
2a^2}$}.\protect\stepcounter{CElett}\addtocounter{equation}{-1}%
\protect\label{eq:SLAC-Infinite-2-n}\setlength{\Cscr}{\value{CEh%
t}\Ctenthex}\addtolength{\Cscr}{-1.0ex}\protect\raisebox{0ex}[%
\value{CEht}\Ctenthex][\Cscr]{}\protect\end{eqnarray}\setcounter
{CEht}{10}\renewcommand\theequation{\arabic{equation}}\par
\refstepcounter{section}\vspace{1.5\baselineskip}\par{\centering
\bf\thesection. Some example applications of the SLAC derivative 
operators\\*[0.5\baselineskip]}\protect\indent\label{sect:Exampl%
esSLAC}To appreciate the behaviour\ of the SLAC derivative opera%
tors it will suffice to examine just a few examples. For conveni%
ence I will here demonstrate some results for one-dimensional la%
ttices with a very small number of sites, as these are sufficien%
t to set the general trend. It is straightforward to verify that 
the same phenomena occur for lattices with a greater number of s%
ites.\par Let us start with the case of \raisebox{0ex}[1ex][0ex]%
{$\protect\displaystyle N=7$}. {}From\ (\protect\ref{eq:SLAC-SLA%
C-1-0}) and (\protect\ref{eq:SLAC-SLAC-1-n-odd}), the first-deri%
vative co\-efficients are just \setcounter{Ceqindent}{0}\protect
\begin{eqnarray}\protect\left.\protect\begin{array}{rcl}\protect
\displaystyle\hspace{-1.3ex}&\protect\displaystyle{}^{(1)\!}c_n^%
{N\:\!\!=7}=\{-0.460,+0.574,-1.034,\,\,0,\,+1.034,-0.574,+0.460%
\},\setlength{\Cscr}{\value{CEht}\Ctenthex}\addtolength{\Cscr}{-%
1.0ex}\protect\raisebox{0ex}[\value{CEht}\Ctenthex][\Cscr]{}%
\protect\end{array}\protect\right.\protect\label{eq:ExamplesSLAC%
-SLAC-1-7}\protect\end{eqnarray}\setcounter{CEht}{10}where the i%
ndex $n$ runs from $-3$ to $+3$, and where for simplicity I have 
set the lattice spacing to \raisebox{0ex}[1ex][0ex]{$\protect
\displaystyle a=1$}. (The co\-efficients will always be rounded 
off to some convenient number of significant figures.) The (disc%
rete) Fourier transform of this operator is just \setcounter{Ceq%
indent}{0}\protect\begin{eqnarray}\protect\left.\protect\begin{a%
rray}{rcl}\protect\displaystyle\hspace{-1.3ex}&\protect
\displaystyle{}^{(1)\!}C_b^{N\:\!\!=7}=\{-i^{\,}1.018,-i^{\,}0.6%
79,-i^{\,}0.339,\,\,0,\,+i^{\,}0.339,+i^{\,}0.679,+i^{\,}1.018\}%
,\setlength{\Cscr}{\value{CEht}\Ctenthex}\addtolength{\Cscr}{-1.%
0ex}\protect\raisebox{0ex}[\value{CEht}\Ctenthex][\Cscr]{}%
\protect\end{array}\protect\right.\protect\label{eq:ExamplesSLAC%
-Fourier-1-7}\protect\end{eqnarray}\setcounter{CEht}{10}where th%
e momentum-space index $b$ also runs from $-3$ to $+3$, where 
\setcounter{Ceqindent}{0}\protect\begin{eqnarray}\protect\left.%
\protect\begin{array}{rcl}\protect\displaystyle\hspace{-1.3ex}&%
\protect\displaystyle p\equiv\mbox{$\protect\displaystyle\protect
\frac{2\pi b}{N^{\!}a}$}.\setlength{\Cscr}{\value{CEht}\Ctenthex
}\addtolength{\Cscr}{-1.0ex}\protect\raisebox{0ex}[\value{CEht}%
\Ctenthex][\Cscr]{}\protect\end{array}\protect\right.\protect
\label{eq:ExamplesSLAC-DefineB}\protect\end{eqnarray}\setcounter
{CEht}{10}It is clear that the Fourier transform (\protect\ref{e%
q:ExamplesSLAC-Fourier-1-7}) has the expected form $ip$, once we 
take account of the factor \raisebox{0ex}[1ex][0ex]{$\protect
\displaystyle1/\sqrt N$} in the definition of the (symmetrical) 
discrete Fourier transform, and note that \setcounter{Ceqindent}%
{0}\protect\begin{eqnarray}\hspace{-1.3ex}&\displaystyle\mbox{$%
\protect\displaystyle\protect\frac{2\pi}{7^{\:\!3/2}}$}\approx0.%
339.\protect\nonumber\setlength{\Cscr}{\value{CEht}\Ctenthex}%
\addtolength{\Cscr}{-1.0ex}\protect\raisebox{0ex}[\value{CEht}%
\Ctenthex][\Cscr]{}\protect\end{eqnarray}\setcounter{CEht}{10}(T%
here is a factor of $\sqrt N$ that floats around all discrete Fo%
urier transform expressions, which would disappear if we more ca%
refully considered how the lattice spacing $a$ should vary when 
the number $N$ of sites is changed, but for our current purposes 
I will simply insert this factor as required; it has no fundamen%
tal significance for the arguments of this paper.)\par If we now 
convolve the operator (\protect\ref{eq:ExamplesSLAC-SLAC-1-7}) w%
ith itself---namely, apply the first-derivative operator twice%
---then we would expect, from the arguments given in \mbox{Sec.~%
$\:\!\!$}\protect\ref{sect:ExpectedSLAC}, that we would obtain t%
he second-order SLAC derivative operator. Performing this convol%
ution numerically, we obtain \setcounter{Ceqindent}{0}\protect
\begin{eqnarray}\protect\left.\protect\begin{array}{rcl}\protect
\displaystyle\hspace{-1.3ex}&\protect\displaystyle\protect\left(%
{}^{(1)\!}c^{N\:\!\!=7}\!\ast\!{}^{(1)\!}c^{N\:\!\!=7}\protect
\right)_{\!\:\!\!n}\!=\{+0.00943,-0.411,+1.928,-3.223,+1.928,-0.%
411,+0.00943\},\setlength{\Cscr}{\value{CEht}\Ctenthex}%
\addtolength{\Cscr}{-1.0ex}\protect\raisebox{0ex}[\value{CEht}%
\Ctenthex][\Cscr]{}\protect\end{array}\protect\right.\protect
\label{eq:ExamplesSLAC-SLAC-1-7-twice}\protect\end{eqnarray}%
\setcounter{CEht}{10}the Fourier transform of which is 
\setcounter{Ceqindent}{0}\protect\begin{eqnarray}\protect\left.%
\protect\begin{array}{rcl}\protect\displaystyle\hspace{-1.3ex}&%
\protect\displaystyle\protect\left({}^{(1)\!}C_b^{N\:\!\!=7}%
\protect\right)^{\!2}\!=\{-2.741,-1.218,-0.305,\,\,0,\,-0.305,-1%
.218,-2.741\},\setlength{\Cscr}{\value{CEht}\Ctenthex}%
\addtolength{\Cscr}{-1.0ex}\protect\raisebox{0ex}[\value{CEht}%
\Ctenthex][\Cscr]{}\protect\end{array}\protect\right.\protect
\label{eq:ExamplesSLAC-Fourier-1-7-twice}\protect\end{eqnarray}%
\setcounter{CEht}{10}which can be recognised to be simply 
\raisebox{0ex}[1ex][0ex]{$\protect\displaystyle-{p}{}^{\raisebox
{-0.25ex}{$\scriptstyle\:\!2$}}$} by noting that \setcounter{Ceq%
indent}{0}\protect\begin{eqnarray}\hspace{-1.3ex}&\displaystyle
\mbox{$\protect\displaystyle\protect\frac{(2\pi)^2}{7^{\:\!5/2}}%
$}\approx0.305.\protect\nonumber\setlength{\Cscr}{\value{CEht}%
\Ctenthex}\addtolength{\Cscr}{-1.0ex}\protect\raisebox{0ex}[%
\value{CEht}\Ctenthex][\Cscr]{}\protect\end{eqnarray}\setcounter
{CEht}{10}It is readily verified that the co\-efficients (%
\protect\ref{eq:ExamplesSLAC-SLAC-1-7-twice}) are precisely thos%
e specified in expressions (\protect\ref{eq:SLAC-SLAC-2-0-odd}) 
and (\protect\ref{eq:SLAC-SLAC-2-n-odd}) for the second-order SL%
AC derivative, as expected.\par Let us now turn to the case of 
\raisebox{0ex}[1ex][0ex]{$\protect\displaystyle N=8$}. We immedi%
ately face a minor source of ambiguity: in the derivative co\-ef%
ficients, should the distances run from \raisebox{0ex}[1ex][0ex]%
{$\protect\displaystyle n=-4$} to \raisebox{0ex}[1ex][0ex]{$%
\protect\displaystyle n=+3$}, or from \raisebox{0ex}[1ex][0ex]{$%
\protect\displaystyle n=-3$} to \raisebox{0ex}[1ex][0ex]{$%
\protect\displaystyle n=+4$}? The case of even $N$ differs from 
that of odd $N$ in that there is a lattice site exactly ``half a 
lattice away'' from the site at which we wish to compute the der%
ivative, and we have to decide whether we ``get to'' this site m%
eans of a distance of \raisebox{0ex}[1ex][0ex]{$\protect
\displaystyle n=-N/2$} or of \raisebox{0ex}[1ex][0ex]{$\protect
\displaystyle n=+N/2$}. Of course, the expression (\protect\ref{%
eq:SLAC-SLAC-1-n-even}) shows us that there will be no contribut%
ion from this site anyway, because for \raisebox{0ex}[1ex][0ex]{%
$\protect\displaystyle n=\pm N/2$} we see from (\protect\ref{eq:%
SLAC-Alpha}) that \raisebox{0ex}[1ex][0ex]{$\protect\displaystyle
\alpha=\pm\pi/2$}, and the co\-efficient (\protect\ref{eq:SLAC-S%
LAC-1-n-even}) is thus proportional to \raisebox{0ex}[1ex][0ex]{%
$\protect\displaystyle\cot(\pm\pi/2)=0$}. (When we ``wrap'' the 
operator around our finite lattice, every contribution at this 
``half-lattice'' position for a positive value of $n$ is cancell%
ed out by an equal and opposite contribution from $-n$.) However%
, it is worth keeping this seemingly trivial annoyance in mind i%
n what follows.\par Arbitrarily choosing the lattice to run from 
\raisebox{0ex}[1ex][0ex]{$\protect\displaystyle n=-3$} to 
\raisebox{0ex}[1ex][0ex]{$\protect\displaystyle n=+4$}, then, th%
e co\-efficients specified by (\protect\ref{eq:SLAC-SLAC-1-0}) a%
nd (\protect\ref{eq:SLAC-SLAC-1-n-even}) for the first-derivativ%
e operation are just \setcounter{Ceqindent}{0}\protect\begin{eqn%
array}\protect\left.\protect\begin{array}{rcl}\protect
\displaystyle\hspace{-1.3ex}&\protect\displaystyle{}^{(1)\!}c_n^%
{N\:\!\!=8}=\{-0.163,+0.393,-0.948,\,\,0,\,+0.948,-0.393,+0.163,%
\,\,0\}.\setlength{\Cscr}{\value{CEht}\Ctenthex}\addtolength{%
\Cscr}{-1.0ex}\protect\raisebox{0ex}[\value{CEht}\Ctenthex][\Cscr
]{}\protect\end{array}\protect\right.\protect\label{eq:ExamplesS%
LAC-SLAC-1-8}\protect\end{eqnarray}\setcounter{CEht}{10}Let us n%
ow immediately apply this first-derivative operator a second tim%
e. When we do so, numerically, we obtain the following co\-effic%
ients: \setcounter{Ceqindent}{0}\protect\begin{eqnarray}\protect
\left.\protect\begin{array}{rcl}\protect\displaystyle\hspace{-1.%
3ex}&\protect\displaystyle\!\!\!\!\protect\left({}^{(1)\!}c^{N\:%
\!\!=8}\!\ast\!{}^{(1)\!}c^{N\:\!\!=8}\protect\right)_{\!\:\!\!n%
}\!=\{-0.872,+0.617,+0.872,-2.159,+0.872,+0.617,-0.872,+0.925\}.%
\!\!\!\setlength{\Cscr}{\value{CEht}\Ctenthex}\addtolength{\Cscr
}{-1.0ex}\protect\raisebox{0ex}[\value{CEht}\Ctenthex][\Cscr]{}%
\protect\end{array}\protect\right.\protect\label{eq:ExamplesSLAC%
-SLAC-1-8-twice}\protect\end{eqnarray}\setcounter{CEht}{10}Somet%
hing is obviously wrong here. The \raisebox{0ex}[1ex][0ex]{$%
\protect\displaystyle n=0$} co\-efficient is negative, as expect%
ed, and the \raisebox{0ex}[1ex][0ex]{$\protect\displaystyle n=\pm
1$} co\-efficients are positive, again as expected, but then the 
\raisebox{0ex}[1ex][0ex]{$\protect\displaystyle n=\pm2$} co\-eff%
icients are \mbox{}\protect\/{\protect\em also\protect\/} positi%
ve, which should not occur. Indeed, the \raisebox{0ex}[1ex][0ex]%
{$\protect\displaystyle n=\pm3$} co\-efficients also have the wr%
ong sign, as does the \raisebox{0ex}[1ex][0ex]{$\protect
\displaystyle n=+4$} co\-efficient.\par Where have we gone wrong%
? If we use expressions (\protect\ref{eq:SLAC-SLAC-2-0-even}) an%
d (\protect\ref{eq:SLAC-SLAC-2-n-even}) to compute the second-or%
der derivative co\-efficients directly, we get a completely diff%
erent set of results: \setcounter{Ceqindent}{0}\protect\begin{eq%
narray}\protect\left.\protect\begin{array}{rcl}\protect
\displaystyle\hspace{-1.3ex}&\protect\displaystyle{}^{(2)\!}c_n^%
{N\:\!\!=8}=\{+0.361,-0.617,+2.106,-3.393,+2.106,-0.617,+0.361,-%
0.308\}.\setlength{\Cscr}{\value{CEht}\Ctenthex}\addtolength{%
\Cscr}{-1.0ex}\protect\raisebox{0ex}[\value{CEht}\Ctenthex][\Cscr
]{}\protect\end{array}\protect\right.\protect\label{eq:ExamplesS%
LAC-SLAC-2-8}\protect\end{eqnarray}\setcounter{CEht}{10}The most 
obvious explanations for this discrepancy are that either there 
is an error in the computational lattice machinery employed, or 
that an algebraic slip has caused errors in (\protect\ref{eq:SLA%
C-SLAC-1-n-even}), (\protect\ref{eq:SLAC-SLAC-2-0-even})\ or (%
\protect\ref{eq:SLAC-SLAC-2-n-even}). However, it can be verifie%
d that neither of these explanations holds water.\par An alterna%
tive way to check the accuracy of (\protect\ref{eq:SLAC-SLAC-1-n%
-even}), (\protect\ref{eq:SLAC-SLAC-2-0-even})\ and (\protect\ref
{eq:SLAC-SLAC-2-n-even}) is to perform the Fourier transform of 
each operator, because we know that the SLAC first- and second-d%
erivative operators should have momentum-space representations o%
f $ip$ and \raisebox{0ex}[1ex][0ex]{$\protect\displaystyle-{p}{}%
^{\raisebox{-0.25ex}{$\scriptstyle\:\!2\!$}}$} respectively---pr%
operties that we have verified for the case of odd~$N$. Now, the 
 Fourier transform of (\protect\ref{eq:ExamplesSLAC-SLAC-1-8}) i%
s \setcounter{Ceqindent}{0}\protect\begin{eqnarray}\protect\left
.\protect\begin{array}{rcl}\protect\displaystyle\hspace{-1.3ex}&%
\protect\displaystyle{}^{(1)\!}C_b^{N\:\!\!=8}=\{-i^{\,}0.833,-i%
^{\,}0.555,-i^{\,}0.278,\,\,0,\,+i^{\,}0.278,+i^{\,}0.555,+i^{\,%
}0.833,\,\,0\,\},\setlength{\Cscr}{\value{CEht}\Ctenthex}%
\addtolength{\Cscr}{-1.0ex}\protect\raisebox{0ex}[\value{CEht}%
\Ctenthex][\Cscr]{}\protect\end{array}\protect\right.\protect
\label{eq:ExamplesSLAC-Fourier-1-8}\protect\end{eqnarray}%
\setcounter{CEht}{10}which is recognised to be $ip$, as expected%
, when we note that \setcounter{Ceqindent}{0}\protect\begin{eqna%
rray}\hspace{-1.3ex}&\displaystyle\mbox{$\protect\displaystyle
\protect\frac{2\pi}{8^{\:\!3/2}}$}\approx0.278.\protect\nonumber
\setlength{\Cscr}{\value{CEht}\Ctenthex}\addtolength{\Cscr}{-1.0%
ex}\protect\raisebox{0ex}[\value{CEht}\Ctenthex][\Cscr]{}\protect
\end{eqnarray}\setcounter{CEht}{10}Likewise, when we take the Fo%
urier transform of the co\-efficients (\protect\ref{eq:ExamplesS%
LAC-SLAC-1-8-twice}), we obtain \setcounter{Ceqindent}{0}\protect
\begin{eqnarray}\protect\left.\protect\begin{array}{rcl}\protect
\displaystyle\hspace{-1.3ex}&\protect\displaystyle\protect\left(%
{}^{(1)\!}C_b^{N\:\!\!=8}\protect\right)^{\!2}\!=\{-1.963,-0.872%
,-0.218,\,\,0,\,-0.218,-0.872,-1.963,\,\,0\,\},\setlength{\Cscr}%
{\value{CEht}\Ctenthex}\addtolength{\Cscr}{-1.0ex}\protect
\raisebox{0ex}[\value{CEht}\Ctenthex][\Cscr]{}\protect\end{array%
}\protect\right.\protect\label{eq:ExamplesSLAC-Fourier-1-8-twice%
}\protect\end{eqnarray}\setcounter{CEht}{10}which is again recog%
nised to be of the expected form \raisebox{0ex}[1ex][0ex]{$%
\protect\displaystyle-{p}{}^{\raisebox{-0.25ex}{$\scriptstyle\:%
\!2\!$}}$} when we note that \setcounter{Ceqindent}{0}\protect
\begin{eqnarray}\hspace{-1.3ex}&\displaystyle\mbox{$\protect
\displaystyle\protect\frac{(2\pi)^2}{8^{\:\!5/2}}$}\approx0.218.%
\protect\nonumber\setlength{\Cscr}{\value{CEht}\Ctenthex}%
\addtolength{\Cscr}{-1.0ex}\protect\raisebox{0ex}[\value{CEht}%
\Ctenthex][\Cscr]{}\protect\end{eqnarray}\setcounter{CEht}{10}Th%
is confirms that the expressions (\protect\ref{eq:SLAC-SLAC-1-n-%
even}), (\protect\ref{eq:SLAC-SLAC-2-0-even})\ and (\protect\ref
{eq:SLAC-SLAC-2-n-even}) are correct, that the convolution is be%
ing performed correctly, and that the Fourier transform engine i%
s also operating correctly.\par So why do the co\-efficients (%
\protect\ref{eq:ExamplesSLAC-SLAC-1-8-twice}) look nothing at al%
l like the expected co\-efficients (\protect\ref{eq:ExamplesSLAC%
-SLAC-2-8})?\par\refstepcounter{section}\vspace{1.5\baselineskip
}\par{\centering\bf\thesection. Why do even lattices give us a p%
roblem?\\*[0.5\baselineskip]}\protect\indent\label{sect:Problem}%
To find out why the even-lattice operators aren't behaving as ex%
pected, we need simply take the Fourier transform of the SLAC se%
cond-derivative co\-efficients (\protect\ref{eq:ExamplesSLAC-SLA%
C-2-8}): \setcounter{Ceqindent}{0}\protect\begin{eqnarray}%
\protect\left.\protect\begin{array}{rcl}\protect\displaystyle
\hspace{-1.3ex}&\protect\displaystyle{}^{(2)\!}C_b^{N\:\!\!=8}=%
\{-1.963,-0.872,-0.218,\,\,0,\,-0.218,-0.872,-1.963,-3.489\}.%
\setlength{\Cscr}{\value{CEht}\Ctenthex}\addtolength{\Cscr}{-1.0%
ex}\protect\raisebox{0ex}[\value{CEht}\Ctenthex][\Cscr]{}\protect
\end{array}\protect\right.\protect\label{eq:Problem-Fourier-2-8}%
\protect\end{eqnarray}\setcounter{CEht}{10}At first glance, this 
appears to be identical to the Fourier transform (\protect\ref{e%
q:ExamplesSLAC-Fourier-1-8-twice}) of the result of the first-de%
rivative operator being applied twice---but there is a crucial d%
ifference: the \raisebox{0ex}[1ex][0ex]{$\protect\displaystyle b%
=+4$} value is $-3.489$ in (\protect\ref{eq:Problem-Fourier-2-8}%
), whereas it \mbox{}\protect\/{\protect\em vanishes\protect\/} 
in (\protect\ref{eq:ExamplesSLAC-Fourier-1-8-twice}).\par What i%
s going on here? The \raisebox{0ex}[1ex][0ex]{$\protect
\displaystyle b=+4$} lattice site in momentum space is, from 
\mbox{Eq.~\raisebox{0ex}[1ex][0ex]{$\protect\displaystyle\!$}}(%
\protect\ref{eq:ExamplesSLAC-DefineB}), simply the site correspo%
nding to the momentum value \raisebox{0ex}[1ex][0ex]{$\protect
\displaystyle p=+\pi/a$}, namely, the very boundary of the first 
Brillouin zone. Now, \mbox{}\protect\/{\protect\em should\protect
\/} the second-derivative operator take a value corresponding to 
\raisebox{0ex}[1ex][0ex]{$\protect\displaystyle-{p}{}^{\raisebox
{-0.25ex}{$\scriptstyle\:\!2\!$}}$} at this boundary site, or sh%
ould it vanish? Obviously, we would prefer it to take its correc%
t (continuum) value of \raisebox{0ex}[1ex][0ex]{$\protect
\displaystyle-{p}{}^{\raisebox{-0.25ex}{$\scriptstyle\:\!2\!$}}$%
} if at all possible; and that is what has been built into the S%
LAC second-derivative operator, which has yielded precisely this 
desired value in (\protect\ref{eq:Problem-Fourier-2-8}). So wher%
e is the problem?\par The problem is not with the second-derivat%
ive operator at all, but rather with our attempt to apply the fi%
rst-derivative operator twice. Applying a derivative operator tw%
ice (\mbox{}\protect\/{\protect\em i.e.\protect\/}, effectively 
convolving the operator with itself) is equivalent to squaring i%
ts momentum-space representation. And so, indeed, the momentum-s%
pace co\-efficients (\protect\ref{eq:ExamplesSLAC-Fourier-1-8-tw%
ice}) are just the squares of the corresponding co\-efficients (%
\protect\ref{eq:ExamplesSLAC-Fourier-1-8}) (when a factor of 
\raisebox{0ex}[1ex][0ex]{$\protect\displaystyle\sqrt N=\sqrt8$} 
is taken into account).\par The problem, however, arises because 
of the momentum-space lattice site at the Brillouin zone boundar%
y, namely, \raisebox{0ex}[1ex][0ex]{$\protect\displaystyle b=+4$%
}. Recall our earlier discussion of the ambiguity arising from t%
he need to decide whether the ``half-lattice distance'' in posit%
ion space should be designated \raisebox{0ex}[1ex][0ex]{$\protect
\displaystyle n=+4$} or \raisebox{0ex}[1ex][0ex]{$\protect
\displaystyle n=-4$}? Here we have the completely analogous ambi%
guity of needing to decide whether the Brillouin zone boundary s%
hould be represented in momentum space by \raisebox{0ex}[1ex][0e%
x]{$\protect\displaystyle b=+4$} or \raisebox{0ex}[1ex][0ex]{$%
\protect\displaystyle b=-4$}. For the SLAC \mbox{}\protect\/{%
\protect\em second\protect\/}-derivative operator, this choice i%
s of no fundamental import, because its Fourier transform, namel%
y, \raisebox{0ex}[1ex][0ex]{$\protect\displaystyle-{p}{}^{%
\raisebox{-0.25ex}{$\scriptstyle\:\!2\!$}}$}, is an even functio%
n of $p$ (and hence $b$); in other words, $(+4)^2$ and $(-4)^2$ 
will both give us the same value for the \raisebox{0ex}[1ex][0ex%
]{$\protect\displaystyle b=\pm4$} site. Indeed, if we were to ar%
bitrarily choose to put the Brillouin zone boundary site at 
\raisebox{0ex}[1ex][0ex]{$\protect\displaystyle b=-4$} in our Fo%
urier transform engine, the Fourier transform of the SLAC second%
-derivative operator (\protect\ref{eq:ExamplesSLAC-SLAC-2-8}) si%
mply becomes \setcounter{Ceqindent}{0}\protect\begin{eqnarray}%
\protect\left.\protect\begin{array}{rcl}\protect\displaystyle
\hspace{-1.3ex}&\protect\displaystyle{}^{(2)\!}C_b^{N\:\!\!=8}=%
\{-3.489,-1.963,-0.872,-0.218,\,\,0,\,-0.218,-0.872,-1.963\}.%
\setlength{\Cscr}{\value{CEht}\Ctenthex}\addtolength{\Cscr}{-1.0%
ex}\protect\raisebox{0ex}[\value{CEht}\Ctenthex][\Cscr]{}\protect
\end{array}\protect\right.\protect\label{eq:Problem-Fourier-2-8-%
negative-hang}\protect\end{eqnarray}\setcounter{CEht}{10}\par Th%
e discrepancy between (\protect\ref{eq:Problem-Fourier-2-8}) and 
(\protect\ref{eq:ExamplesSLAC-Fourier-1-8-twice}) tells us that 
it is the ``missing'' value at \raisebox{0ex}[1ex][0ex]{$\protect
\displaystyle b=+4$} in (\protect\ref{eq:ExamplesSLAC-Fourier-1-%
8-twice}) that is responsible for us not obtaining the expected 
result when we applied the SLAC first-derivative operator twice. 
In isolation, this seems reasonable enough: we simply need to ma%
ke sure that the ``missing'' value is put back in. But things st%
art to get rather strange when we investigate the ramifications 
of this requirement. Presumably, we must ``repair'' the \mbox{}%
\protect\/{\protect\em first\protect\/}-derivative operator, to 
ensure that, when we apply it twice (and hence square its repres%
entation in momentum space), the ``missing'' value is restored. 
This immediately draws our attention to the final co\-efficient 
in (\protect\ref{eq:ExamplesSLAC-Fourier-1-8}), which currently 
vanishes. The fundamental reason that it \mbox{}\protect\/{%
\protect\em must\protect\/} vanish is not difficult to appreciat%
e: the momentum eigenstate at the Brillouin zone boundary oscill%
ates in position space at the maximum possible (Nyquist) frequen%
cy, \raisebox{0ex}[1ex][0ex]{$\protect\displaystyle+1,-1,+1,-1,%
\ldots\,$}, which is an even function of $n$. Now, the SLAC firs%
t-derivative operator is an odd function of $n$, and furthermore 
vanishes at \raisebox{0ex}[1ex][0ex]{$\protect\displaystyle n=+N%
/2$} (or \raisebox{0ex}[1ex][0ex]{$\protect\displaystyle n=-N/2$%
} if that had been our arbitrary choice), so no matter which pos%
ition we ``place'' it next to the maximum-frequency eigenstate, 
in the convolution, it will necessarily yield a vanishing sum. F%
or example, with \raisebox{0ex}[1ex][0ex]{$\protect\displaystyle
N=8$}, the operator (\protect\ref{eq:ExamplesSLAC-SLAC-1-8}) wil%
l lead to a convolution of the form \setcounter{Ceqindent}{0}%
\protect\begin{eqnarray}\hspace{-1.3ex}&\displaystyle-0.163-0.39%
3-0.948+0+0.948+0.393+0.163+0=0,\protect\nonumber\setlength{\Cscr
}{\value{CEht}\Ctenthex}\addtolength{\Cscr}{-1.0ex}\protect
\raisebox{0ex}[\value{CEht}\Ctenthex][\Cscr]{}\protect\end{eqnar%
ray}\setcounter{CEht}{10}or its negative, for all positions. Thi%
s symmetry argument, in fact, proves the following general theor%
em: \mbox{}\protect\/{\protect\em Any position-space operator th%
at is an odd function of \protect\/}$n$\mbox{}\protect\/{\protect
\em, and vanishes at \protect\/}\raisebox{0ex}[1ex][0ex]{$%
\protect\displaystyle n=\pm N/2$}\mbox{}\protect\/{\protect\em, 
will lead to a zero at the Brillouin zone boundary for a lattice 
with an even number of sites\protect\/}. (This is, of course, al%
so true for the na\-\"\i ve\ first-derivative operator~[\ref{cit%
:Costella2002a}], although in that particular case the restricte%
d nature of this theorem---namely, that the necessary vanishing 
of the Fourier transform is restricted to the single site that l%
ies on the Brillouin zone boundary---is somewhat hidden by the f%
act that the Fourier transform of the na\-\"\i ve\ operator appr%
oaches this zero smoothly, rather than abruptly as is the case h%
ere.)\par Why does this theorem only hold for lattices with an e%
ven number of sites? Simply because lattices with an odd number 
of sites do not possess a momentum eigenstate that is on the Bri%
llouin zone boundary. Not only is this obvious from the relation 
(\protect\ref{eq:ExamplesSLAC-DefineB})---namely, for \raisebox{%
0ex}[1ex][0ex]{$\protect\displaystyle p=\pm\pi/a$} we require 
\raisebox{0ex}[1ex][0ex]{$\protect\displaystyle b=N/2$}, which i%
s only possible if $N$ is even---but it is also clear in positio%
n space, because such an eigenstate, oscillating \raisebox{0ex}[%
1ex][0ex]{$\protect\displaystyle+1,-1,+1,-1,\ldots\,$}, necessar%
ily requires an even number of sites in which to fit an integral 
number of \raisebox{0ex}[1ex][0ex]{$\protect\displaystyle+1,-1$} 
pairs; otherwise, the periodic boundary conditions of the lattic%
e would not be satisfied (\mbox{}\protect\/{\protect\em i.e.%
\protect\/}, viewed as a ring, there would be either two $+1$ or 
two $-1$ sites next to each other). As should be clear from the 
above example, the general theorem applies \mbox{}\protect\/{%
\protect\em only\protect\/} to the single momentum site at the B%
rillouin zone boundary, so if this site does not exist (as is th%
e case for odd $N$), the theorem likewise fails to exist.\par Th%
is general problem may, in fact, have been the reason why 
\protect\ref{au:Drell1976a} restricted their attention exclusive%
ly to lattices with an odd number of sites---namely, to avoid th%
e conundrum that we are now facing while trying to apply the fir%
st- and second-order SLAC derivative operators. (However, I have 
no direct evidence that this was their motivation for this restr%
iction.)\par\refstepcounter{section}\vspace{1.5\baselineskip}\par
{\centering\bf\thesection. Solving the even-lattice conundrum\\*%
[0.5\baselineskip]}\protect\indent\label{sect:Solution}It is all 
well and good to understand why our even-lattice operators do no%
t have the properties we expect of them, but our ultimate goal m%
ust be to ``repair'' the first-derivative operator so that we ca%
n recover these properties.\par Clearly, we must ensure that the 
second-derivative operator possesses its correct (continuum) val%
ue \raisebox{0ex}[1ex][0ex]{$\protect\displaystyle-{p}{}^{%
\raisebox{-0.25ex}{$\scriptstyle\:\!2\!$}}$} at the Brillouin zo%
ne boundary, as we have found is explicitly obtained if we use t%
he SLAC second-order derivative co\-efficients derived in [\ref{%
cit:Costella2002b}]. To maintain the self-consistency of differe%
ntial calculus on such lattices, we wish to further ensure that 
two successive applications of the first-derivative operator be 
identically equivalent to a single application of the second-der%
ivative operator; if we didn't insist on reinstating this proper%
ty, there would be little point in using the SLAC operators at a%
ll---we may as well go back to using the na\-\"\i ve\ operators, 
which do not obey this property except in the limit of small $p$%
.\par Since a double application of the first-derivative operato%
r is equivalent to squaring its representation in momentum space%
, our task must be to somehow replace the vanishing final co\-ef%
ficient of (\protect\ref{eq:ExamplesSLAC-Fourier-1-8}) with its 
correct (continuum) value of $ip$. In the case of (\protect\ref{%
eq:ExamplesSLAC-Fourier-1-8}) (taking into account the factor of 
\raisebox{0ex}[1ex][0ex]{$\protect\displaystyle2\pi/\!\sqrt8$}), 
we need to modify the momentum-space representation to 
\setcounter{Ceqindent}{0}\protect\begin{eqnarray}\protect\left.%
\protect\begin{array}{rcl}\protect\displaystyle\hspace{-1.3ex}&%
\protect\displaystyle{}^{(1)\!}C_b^{N\:\!\!=8}=\{-i^{\,}0.833,-i%
^{\,}0.555,-i^{\,}0.278,\,\,0,\,+i^{\,}0.278,+i^{\,}0.555,+i^{\,%
}0.833,+i^{\,}1.111\}.\setlength{\Cscr}{\value{CEht}\Ctenthex}%
\addtolength{\Cscr}{-1.0ex}\protect\raisebox{0ex}[\value{CEht}%
\Ctenthex][\Cscr]{}\protect\end{array}\protect\right.\protect
\label{eq:Solution-Fourier-1-8-fixed-positive-hang}\protect\end{%
eqnarray}\setcounter{CEht}{10}More precisely (and taking more ca%
reful note, now, of factors of $a$ and $\sqrt N$), we wish to ad%
d an extra value of \raisebox{0ex}[1ex][0ex]{$\protect
\displaystyle+i\pi/a\sqrt N$} at the momentum value \raisebox{0e%
x}[1ex][0ex]{$\protect\displaystyle p=+\pi/a$}. Taking into acco%
unt the factor of \raisebox{0ex}[1ex][0ex]{$\protect\displaystyle
1/\!\sqrt N$} in the discrete Fourier transform, this means that 
we wish to add to the SLAC first-derivative operator an \mbox{}%
\protect\/{\protect\em extra\protect\/} function in position spa%
ce, of the form \setcounter{Ceqindent}{0}\protect\begin{eqnarray%
}\protect\left.\protect\begin{array}{rcl}\protect\displaystyle
\hspace{-1.3ex}&\protect\displaystyle{}^{(2)\!}c_{\mbox{%
\scriptsize extra}}^{\mbox{\scriptsize$N\!$ even}}=\mbox{$%
\protect\displaystyle\protect\frac{i\pi}{N^{\!}a}$}(-1)^{n\!}.%
\setlength{\Cscr}{\value{CEht}\Ctenthex}\addtolength{\Cscr}{-1.0%
ex}\protect\raisebox{0ex}[\value{CEht}\Ctenthex][\Cscr]{}\protect
\end{array}\protect\right.\protect\label{eq:Solution-Extra-posit%
ive-hang}\protect\end{eqnarray}\setcounter{CEht}{10}In other wor%
ds, for our example of \raisebox{0ex}[1ex][0ex]{$\protect
\displaystyle N=8$}, we wish to change the first-derivative oper%
ator from the co\-efficients listed in (\protect\ref{eq:Examples%
SLAC-SLAC-1-8}) to \setcounter{Ceqindent}{0}\protect\begin{eqnar%
ray}\protect\left.\protect\begin{array}{rcl}\protect\displaystyle
\hspace{-1.3ex}&&\hspace{-1.3ex}\protect\displaystyle{}^{(1)\!}c%
_b^{N\:\!\!=8}=\{\,-0.163-i^{\,}0.393,\;\,+0.393+i^{\,}0.393,\;%
\,-0.948-i^{\,}0.393,\;\,\,\,0+i^{\,}0.393,\setcounter{Ceqindent%
}{120}\setlength{\Cscr}{\value{CEht}\Ctenthex}\addtolength{\Cscr
}{-1.0ex}\protect\raisebox{0ex}[\value{CEht}\Ctenthex][\Cscr]{}%
\\*[0.55ex]\protect\displaystyle\hspace{-1.3ex}&\protect
\displaystyle&\hspace{-1.3ex}\protect\displaystyle{\protect\mbox
{}}\hspace{\value{Ceqindent}\Ctenthex}+0.948-i^{\,}0.393,\;\,-0.%
393+i^{\,}0.393,\;\,+0.163-i^{\,}0.393,\;\,\,\,0+i^{\,}0.393\,\}%
,\setlength{\Cscr}{\value{CEht}\Ctenthex}\addtolength{\Cscr}{-1.%
0ex}\protect\raisebox{0ex}[\value{CEht}\Ctenthex][\Cscr]{}%
\protect\end{array}\protect\right.\protect\label{eq:Solution-SLA%
C-1-8-fixed-positive-hang}\protect\end{eqnarray}\setcounter{CEht%
}{10}where \setcounter{Ceqindent}{0}\protect\begin{eqnarray}%
\hspace{-1.3ex}&\displaystyle\mbox{$\protect\displaystyle\protect
\frac{\pi}{8}$}\approx0.393.\protect\nonumber\setlength{\Cscr}{%
\value{CEht}\Ctenthex}\addtolength{\Cscr}{-1.0ex}\protect
\raisebox{0ex}[\value{CEht}\Ctenthex][\Cscr]{}\protect\end{eqnar%
ray}\setcounter{CEht}{10}Clearly, the extra contribution (%
\protect\ref{eq:Solution-Extra-positive-hang}) to (\protect\ref{%
eq:Solution-SLAC-1-8-fixed-positive-hang}), being an even functi%
on of $n$, avoids the consequences of the general theorem above, 
as it must by the way it has been constructed. And, sure enough, 
if we apply the operator (\protect\ref{eq:Solution-SLAC-1-8-fixe%
d-positive-hang}) twice, we obtain precisely the SLAC second-der%
ivative co\-efficients (\protect\ref{eq:ExamplesSLAC-SLAC-2-8}), 
as expected.\par But what sort of a monster have we created?\par
There are a number of important objections that one may wish to 
make at this point. Most seriously, we made a completely arbitra%
ry choice, when we first considered the Fourier transform to mom%
entum space, to place the Brillouin zone boundary site at 
\raisebox{0ex}[1ex][0ex]{$\protect\displaystyle b=+4$}. What if 
we had, instead, chosen to put it at \raisebox{0ex}[1ex][0ex]{$%
\protect\displaystyle b=-4$}? In the case of the second-derivati%
ve operator, such a change makes nothing but a notational differ%
ence, because the Fourier transform of the second-derivative ope%
rator is an even function of~$b$. In the case of the first-deriv%
ative, on the other hand, we must surely insist that the extra s%
ite take on its correct (continum) value, namely, \raisebox{0ex}%
[1ex][0ex]{$\protect\displaystyle-i^{\,}1.111$}, which is the ne%
gative of that added to (\protect\ref{eq:Solution-Fourier-1-8-fi%
xed-positive-hang}), so that the momentum-space representation b%
ecomes \setcounter{Ceqindent}{0}\protect\begin{eqnarray}\protect
\left.\protect\begin{array}{rcl}\protect\displaystyle\hspace{-1.%
3ex}&\protect\displaystyle{}^{(1)\!}C_b^{N\:\!\!=8}=\{-i^{\,}1.1%
11,-i^{\,}0.833,-i^{\,}0.555,-i^{\,}0.278,\,\,0,\,+i^{\,}0.278,+%
i^{\,}0.555,+i^{\,}0.833\}.\setlength{\Cscr}{\value{CEht}%
\Ctenthex}\addtolength{\Cscr}{-1.0ex}\protect\raisebox{0ex}[%
\value{CEht}\Ctenthex][\Cscr]{}\protect\end{array}\protect\right
.\protect\label{eq:Solution-Fourier-1-8-fixed-negative-hang}%
\protect\end{eqnarray}\setcounter{CEht}{10}Now, since the positi%
on-space representation of a term at the Brillouin zone boundary 
is unchanged, regardless of whether we write it as \raisebox{0ex%
}[1ex][0ex]{$\protect\displaystyle b=+4$} or \raisebox{0ex}[1ex]%
[0ex]{$\protect\displaystyle b=-4$}, because \setcounter{Ceqinde%
nt}{0}\protect\begin{eqnarray}\hspace{-1.3ex}&\displaystyle e^{+%
i\pi n}\equiv(-1)^n\equiv e^{-i\pi n\:\!\!},\protect\nonumber
\setlength{\Cscr}{\value{CEht}\Ctenthex}\addtolength{\Cscr}{-1.0%
ex}\protect\raisebox{0ex}[\value{CEht}\Ctenthex][\Cscr]{}\protect
\end{eqnarray}\setcounter{CEht}{10}this change of sign for the e%
xtra momentum-space term implies that the extra function added i%
n position space will likewise change in sign: \setcounter{Ceqin%
dent}{0}\protect\begin{eqnarray}\protect\left.\protect\begin{arr%
ay}{rcl}\protect\displaystyle\hspace{-1.3ex}&\protect
\displaystyle{}^{(2)\!}c_{\mbox{\scriptsize extra}}^{\mbox{%
\scriptsize$N\!$ even}}=\mbox{$\protect\displaystyle\protect\frac
{i\pi}{N^{\!}a}$}(-1)^{n+1\:\!\!},\setlength{\Cscr}{\value{CEht}%
\Ctenthex}\addtolength{\Cscr}{-1.0ex}\protect\raisebox{0ex}[%
\value{CEht}\Ctenthex][\Cscr]{}\protect\end{array}\protect\right
.\protect\label{eq:Solution-Extra-negative-hang}\protect\end{eqn%
array}\setcounter{CEht}{10}which for our example of \raisebox{0e%
x}[1ex][0ex]{$\protect\displaystyle N=8$} yields \setcounter{Ceq%
indent}{0}\protect\begin{eqnarray}\protect\left.\protect\begin{a%
rray}{rcl}\protect\displaystyle\hspace{-1.3ex}&&\hspace{-1.3ex}%
\protect\displaystyle{}^{(1)\!}c_b^{N\:\!\!=8}=\{\,\,0-i^{\,}0.3%
93,\;\,-0.163+i^{\,}0.393,\;\,+0.393-i^{\,}0.393,\;\,-0.948+i^{%
\,}0.393,\setcounter{Ceqindent}{130}\setlength{\Cscr}{\value{CEh%
t}\Ctenthex}\addtolength{\Cscr}{-1.0ex}\protect\raisebox{0ex}[%
\value{CEht}\Ctenthex][\Cscr]{}\\*[0.55ex]\protect\displaystyle
\hspace{-1.3ex}&\protect\displaystyle&\hspace{-1.3ex}\protect
\displaystyle{\protect\mbox{}}\hspace{\value{Ceqindent}\Ctenthex
}\,\,0-i^{\,}0.393,\;\,+0.948+i^{\,}0.393,\;\,-0.393-i^{\,}0.393%
,\;\,+0.163+i^{\,}0.393\}.\setlength{\Cscr}{\value{CEht}\Ctenthex
}\addtolength{\Cscr}{-1.0ex}\protect\raisebox{0ex}[\value{CEht}%
\Ctenthex][\Cscr]{}\protect\end{array}\protect\right.\protect
\label{eq:Solution-SLAC-1-8-fixed-negative-hang}\protect\end{eqn%
array}\setcounter{CEht}{10}Again, if we apply this operator twic%
e, we obtain the expected analogue\ of (\protect\ref{eq:Examples%
SLAC-SLAC-2-8}) with a site at \raisebox{0ex}[1ex][0ex]{$\protect
\displaystyle n=-4$} rather than \raisebox{0ex}[1ex][0ex]{$%
\protect\displaystyle n=+4$}: \setcounter{Ceqindent}{0}\protect
\begin{eqnarray}\protect\left.\protect\begin{array}{rcl}\protect
\displaystyle\hspace{-1.3ex}&\protect\displaystyle\!\!\!\!\!%
\protect\left({}^{(1)\!}c^{N\:\!\!=8}\!\ast\!{}^{(1)\!}c^{N\:\!%
\!=8}\protect\right)_{\!\:\!\!n}\!=\{-0.308,+0.361,-0.617,+2.106%
,-3.393,+2.106,-0.617,+0.361\}.\!\!\!\!\setlength{\Cscr}{\value{%
CEht}\Ctenthex}\addtolength{\Cscr}{-1.0ex}\protect\raisebox{0ex}%
[\value{CEht}\Ctenthex][\Cscr]{}\protect\end{array}\protect\right
.\protect\label{eq:Solution-SLAC-2-8-negative-hang}\protect\end{%
eqnarray}\setcounter{CEht}{10}In other words, we can add in \mbox
{}\protect\/{\protect\em either\protect\/} of the extra function%
s (\protect\ref{eq:Solution-Extra-positive-hang}) \mbox{}\protect
\/{\protect\em or\protect\/} (\protect\ref{eq:Solution-Extra-neg%
ative-hang})---fundamentally depending only on which end of the 
momentum-space lattice we choose to put the Brillouin zone bound%
ary term---and, either way, we still get the right answers: 
\setcounter{Ceqindent}{0}\protect\begin{eqnarray}\protect\left.%
\protect\begin{array}{rcl}\protect\displaystyle\hspace{-1.3ex}&%
\protect\displaystyle{}^{(2)\!}c_{\mbox{\scriptsize extra}}^{%
\mbox{\scriptsize$N\!$ even}}=\pm\mbox{$\protect\displaystyle
\protect\frac{i\pi}{N^{\!}a}$}(-1)^{n\!}.\setlength{\Cscr}{\value
{CEht}\Ctenthex}\addtolength{\Cscr}{-1.0ex}\protect\raisebox{0ex%
}[\value{CEht}\Ctenthex][\Cscr]{}\protect\end{array}\protect
\right.\protect\label{eq:Solution-Extra-both}\protect\end{eqnarr%
ay}\setcounter{CEht}{10}So which function is ``correct''? Which 
sign is the ``right'' sign: the positive or the negative?\par
\refstepcounter{section}\vspace{1.5\baselineskip}\par{\centering
\bf\thesection. ``Spontaneous symmetry breaking'' for even latti%
ces\\*[0.5\baselineskip]}\protect\indent\label{sect:SSB}The simp%
le answer to the question of which sign is ``correct'' in (%
\protect\ref{eq:Solution-Extra-both}) is that they are \mbox{}%
\protect\/{\protect\em both\protect\/} ``correct''\mbox{$\!$}. W%
hen we create a lattice with an even number of sites, we must de%
cide whether the ``half-lattice'' distance in position space is 
to be located at \raisebox{0ex}[1ex][0ex]{$\protect\displaystyle
n=+N/2$} or \raisebox{0ex}[1ex][0ex]{$\protect\displaystyle n=-N%
/2$}, and we must likewise decided whether the Brillouin zone bo%
undary site in momentum space is to be located at \raisebox{0ex}%
[1ex][0ex]{$\protect\displaystyle b=+N/2$} or \raisebox{0ex}[1ex%
][0ex]{$\protect\displaystyle b=-N/2$}. It is the latter choice 
that determines whether the extra term required in the SLAC firs%
t-derivative operator should come in with the plus sign or the m%
inus sign in (\protect\ref{eq:Solution-Extra-both}).\par Althoug%
h it does not have the same physical and dynamical connotations, 
this situation is---mathematically, at least---completely analog%
ous to the situation of spontaneous symmetry breaking in field t%
heories. Here, the SLAC first-derivative operator is a real, odd 
function of the position-space distance $n$, for the case of an 
infinite lattice, which implies that its momentum-space represen%
tation is a purely imaginary, odd function of $p$. When we ``wra%
pped'' this infinite-lattice operator around a finite lattice, i%
n [\ref{cit:Costella2002b}], these properties were maintained. T%
he consequence, however, was the introduction of a spurious zero 
into its momentum-space representation, at the Brillouin zone bo%
undary. The fundamental problem is that a lattice with an even n%
umber of sites will always have a momentum eigenstate that is ``%
unpaired''; in other words, an even lattice by necessity breaks 
the symmetry in momentum space between positive and negative mom%
enta. Now, if we do not introduce the extra function (\protect
\ref{eq:Solution-Extra-both}) into the first-derivative operator%
, we are effectively imposing the odd, imaginary property in mom%
entum space on the operator by treating the site at the Brilloui%
n zone boundary as being \mbox{}\protect\/{\protect\em both%
\protect\/} \raisebox{0ex}[1ex][0ex]{$\protect\displaystyle p=+%
\pi/a$} \mbox{}\protect\/{\protect\em and\protect\/} \raisebox{0%
ex}[1ex][0ex]{$\protect\displaystyle p=-\pi/a$}; and since, to b%
e an odd function of $p$, these sites should have equal and oppo%
site values, we are effectively ``averaging'' them out to give z%
ero. Put differently, the only way for an imaginary number to be 
its own negative is for the number to be zero.\par The example o%
f spontaneous symmetry breaking in field theory guides us in the 
current case. In field theory, we do not impose the underlying s%
ymmetry on the states of the broken-symmetry formalism; rather, 
we recognise that the symmetry has, indeed, been broken, and we 
embrace and analyse the ramifications of the symmetry being brok%
en---understanding that for every arbitrary choice of manner of 
breaking of the fundamental symmetry, there will be a physically 
equivalent description (or an infinite number of such descriptio%
ns, for a continuous symmetry) with a different arbitrary choice 
of manner of breaking. Analogously, for an even lattice, we can 
choose to put the Brillouin zone boundary site at \mbox{}\protect
\/{\protect\em either\protect\/} \raisebox{0ex}[1ex][0ex]{$%
\protect\displaystyle b=+N/2$} or \raisebox{0ex}[1ex][0ex]{$%
\protect\displaystyle b=-N/2$}, in the full knowledge that this 
choice is arbitrary, but that \mbox{}\protect\/{\protect\em it m%
ust be made for any definite application of the lattice for prac%
tical purposes\protect\/}.\par This is, admittedly, a somewhat 
``philosophical'' justification for why the extra function (%
\protect\ref{eq:Solution-Extra-both}) \mbox{}\protect\/{\protect
\em should\protect\/} be added to the SLAC first-derivative oper%
ator---in other words, that its omission would arguably lead to 
a greater divergence from the continuum reality than its inclusi%
on. The ``proof of the pudding''\mbox{$\!$}, of course, will be 
the application of this prescription to real-world calculations. 
However, as these will take time to implement, it is worth discu%
ssing here some of the more fundamental properties of the modifi%
ed first-derivative operator in general terms, so that we know w%
hat we might expect when it is used in practice.\par
\refstepcounter{section}\vspace{1.5\baselineskip}\par{\centering
\bf\thesection. Properties of the modified SLAC first-derivative 
operator\\*[0.5\baselineskip]}\protect\indent\label{sect:Propert%
ies}Earlier I argued that the most serious objection to the modi%
fication to the SLAC first-derivative operator must be the arbit%
rariness of the choice of sign in (\protect\ref{eq:Solution-Extr%
a-both}). Almost equally objectionable, however, must be the 
\mbox{}\protect\/{\protect\em form\protect\/} of the additional 
function (\protect\ref{eq:Solution-Extra-both}). The derivative 
operator is, naturally, a real operator in position space (as it 
must be, because the process of differentiation is something tha%
t can be applied to real functions, not just complex functions). 
The additional function (\protect\ref{eq:Solution-Extra-both}), 
in contrast, is purely imaginary. Won't this destroy some fundam%
ental property of calculus? Won't it mess up the unitarity of an%
y quantum calculation?\par It is the unitarity of the formalism 
that is of most concern to us here. As reviewed in [\ref{cit:Cos%
tella2002a}], the only reason that the na\-\"\i ve\ first-deriva%
tive operator possesses a spurious zero at the Brillouin zone bo%
undary---leading, fundamentally, to the fermion doubling problem%
---is that unitarity prevents us from computing a first-derivati%
ve on the basis of adjacent lattice sites. And, indeed, unitarit%
y is such an important property of any quantum calculation that 
it would be difficult to proceed with confidence at all if it we%
re broken---regardless of whether or not it would be restored in 
the continuum limit.\par So let us determine whether the extra f%
unction (\protect\ref{eq:Solution-Extra-both}) will cause us any 
problems with unitarity. The way to check this is to examine the 
momentum operator $p$ in position space, namely, $-i$ times the 
derivative operator. When we write this operator as a matrix (wi%
th columns representing the ``input'' function, and rows the ``o%
utput'' function, namely, the derivative of the original functio%
n), we require it to be Hermitian~[\ref{cit:Costella2002a}]. Now%
, (\protect\ref{eq:Solution-Extra-both}) yields \setcounter{Ceqi%
ndent}{0}\protect\begin{eqnarray}\protect\left.\protect\begin{ar%
ray}{rcl}\protect\displaystyle\hspace{-1.3ex}&\protect
\displaystyle p_{\mbox{\scriptsize extra}}^{\mbox{\scriptsize$N%
\!$ even}}\equiv-i\,{}^{(2)\!}c_{\mbox{\scriptsize extra}}^{\mbox
{\scriptsize$N\!$ even}}=\pm\mbox{$\protect\displaystyle\protect
\frac{\pi}{N^{\!}a}$}(-1)^{n\!}.\setlength{\Cscr}{\value{CEht}%
\Ctenthex}\addtolength{\Cscr}{-1.0ex}\protect\raisebox{0ex}[%
\value{CEht}\Ctenthex][\Cscr]{}\protect\end{array}\protect\right
.\protect\label{eq:Properties-P-Extra}\protect\end{eqnarray}%
\setcounter{CEht}{10}This represents a matrix operator of the fo%
rm \setcounter{Ceqindent}{0}\protect\begin{eqnarray}\protect\left
.\protect\begin{array}{rcl}\protect\displaystyle\hspace{-1.3ex}&%
\protect\displaystyle p_{\mbox{\scriptsize extra}}^{\mbox{%
\scriptsize$N\!$ even}}=\pm\mbox{$\protect\displaystyle\protect
\frac{\pi}{N^{\!}a}$}\!\protect\left(\begin{array}{cccccc}+1&-1&%
+1&-1&+1&\cdots\\[0.5ex]-1&+1&-1&+1&-1&\cdots\\[0.5ex]+1&-1&+1&-%
1&+1&\cdots\\[0.5ex]-1&+1&-1&+1&-1&\cdots\\[0.5ex]+1&-1&+1&-1&+1%
&\cdots\\\vdots&\vdots&\vdots&\!\vdots\!&\!\vdots\!&\ddots\end{a%
rray}\protect\right)\!.\setlength{\Cscr}{\value{CEht}\Ctenthex}%
\addtolength{\Cscr}{-1.0ex}\protect\raisebox{0ex}[\value{CEht}%
\Ctenthex][\Cscr]{}\protect\end{array}\protect\right.\protect
\label{eq:Properties-P-extra-matrix}\protect\end{eqnarray}%
\setcounter{CEht}{10}Although rather unfamiliar as an operator i%
n everyday calculations, this matrix most surely is Hermitian. T%
hus, we cannot object to the extra function (\protect\ref{eq:Sol%
ution-Extra-both}) on the grounds of unitarity: its inclusion wi%
ll not affect the unitarity of our calculations in the least.\par
Of course, if we were looking to use an even lattice to perform 
some sort of calculation of strictly real functions (such as for 
an Engineering application, say), then we would still be able to 
object to (\protect\ref{eq:Solution-Extra-both}) on the grounds 
that it would introduce ``spurious'' imaginary components into t%
he calculations. On the other hand, \mbox{}\protect\/{\protect\em
any\protect\/} calculation on a lattice is somewhat artificial a%
nd mathematical anyway---and if the introduction of imaginary co%
mponents is the price one has to pay to ensure that the self-con%
sistency of differential calculus is maintained, on such a latti%
ce, then it is arguable that such an introduction should indeed 
be made, regardless of the nature of the application.\par Howeve%
r, it is worth examining how large in magnitude the extra terms 
(\protect\ref{eq:Solution-Extra-both}) in the differential opera%
tor will actually be, relative to the ``normal'' terms (\protect
\ref{eq:SLAC-SLAC-1-n-even}). Apart from a numerical factor of $%
\pi$, the magnitude of each extra term (\protect\ref{eq:Solution%
-Extra-both}) is a factor of $n/N$ smaller than the correspondin%
g term in (\protect\ref{eq:SLAC-SLAC-1-n-even}). Thus, the ``ext%
ra'' terms are of roughly the same size as the ``normal'' terms 
at the farthest distances on the periodic lattice.\par What does 
this mean? Remember that, in principle, we are actually trying t%
o take a spatial derivative of a function \mbox{}\protect\/{%
\protect\em locally\protect\/}, namely, that the derivative func%
tion at any position $x$ should only depend on that part of the 
original function that is ``local'' to $x$. The SLAC derivative 
operators appear to violate this wish, because the derivative at 
any lattice site $n$ depends on the value of the original functi%
on at \mbox{}\protect\/{\protect\em all\protect\/} sites on the 
lattice. However, as emphasised in [\ref{cit:Costella2002a}], th%
is ``non-locality'' is an illusion, because as the number of sit%
es $N$ is increased, the lattice spacing is decreased, so that t%
he contribution to the SLAC derivative operator from any finite 
(real) distance away from the point of differentiation vanishes 
as $N$ is increased. Likewise, as $N$ is increased, the relative 
contribution of the extra function (\protect\ref{eq:Solution-Ext%
ra-both}) is decreased, and in the limit \raisebox{0ex}[1ex][0ex%
]{$\protect\displaystyle N\rightarrow\infty$} it vanishes.\par A 
further property of the modified derivative operator is obvious 
from the way we have constructed it, namely, that the only zero 
in its momentum-space representation is at \raisebox{0ex}[1ex][0%
ex]{$\protect\displaystyle p=0$}; we have eliminated the spuriou%
s zero at the Brillouin zone boundary. Again, this simply restor%
es us to the situation we already have for the case of odd $N$, 
namely, that there is no spurious zero to start with for odd $N$%
, because there is no momentum-space lattice site on the Brillou%
in zone boundary. One would expect that this ``momentum space fi%
delity'' will ensure that there is no problem with fermion doubl%
ing in lattice field theory, as there is for the na\-\"\i ve\ fi%
rst-derivative operator.\par\refstepcounter{section}\vspace{1.5%
\baselineskip}\par{\centering\bf\thesection. Conclusions\\*[0.5%
\baselineskip]}\protect\indent\label{sect:Conclusions}In this pa%
per I have tried to show why lattices with an even number of sit%
es contain potential pitfalls for the unwary. I have illustrated 
this with the example of my own blundering attempts to correctly 
define and implement the SLAC derivative operators on simple one%
-dimensional lattices. However, the fundamental principle---and 
warning---is, I believe, a general one; similar subtleties shoul%
d apply to other aspects of field theory formulated on even latt%
ices. The fundamental reason for this, I believe, is that such l%
attices break the symmetry between positive and negative momentu%
m values that holds for any continuum formalism (and which is re%
tained for odd lattices); a faithful representation on an even l%
attice requires us to ``spontaneously break'' this symmetry in f%
avour\ of an extra momentum eigenstate at the Brillouin zone bou%
ndary of either \raisebox{0ex}[1ex][0ex]{$\protect\displaystyle
p=+\pi/a$} or \raisebox{0ex}[1ex][0ex]{$\protect\displaystyle p=%
-\pi/a$}, but not both.\par Given the computational convenience 
of employing lattices with an even number of sites (generally a 
power of two in each dimension), the full exploration of these s%
ubtleties may prove to be of interest to practitioners in the fi%
eld of lattice calculations. I will be continuing these investig%
ations in due course, but at present the full potential ramifica%
tions are not clear to me.\par\vspace{1.5\baselineskip}\par{%
\centering\bf Acknowledgments\\*[0.5\baselineskip]}\protect
\indent Many helpful discussions with Tien D.\ Kieu are grateful%
ly acknowledged.\par\vspace{1.5\baselineskip}\par{\centering\bf
References\\*[0.5\baselineskip]}{\protect\mbox{}}\vspace{-%
\baselineskip}\vspace{-2ex}\settowidth\CGDnum{[\ref{citlast}]}%
\setlength{\CGDtext}{\textwidth}\addtolength{\CGDtext}{-\CGDnum}%
\begin{list}{Error!}{\setlength{\labelwidth}{\CGDnum}\setlength{%
\labelsep}{0.75ex}\setlength{\leftmargin}{\labelwidth}%
\addtolength{\leftmargin}{\labelsep}\setlength{\rightmargin}{0ex%
}\setlength{\itemsep}{0ex}\setlength{\parsep}{0ex}}\protect
\frenchspacing\setcounter{CBtnc}{1}\addtocounter{CBcit}{1}\item[%
\hfill{[}\arabic{CBcit}{]}]\renewcommand\theCscr{\arabic{CBcit}}%
\protect\refstepcounter{Cscr}\protect\label{cit:Costella2002a}J.%
 ~P.~Costella, \renewcommand\theCscr{Costella}\protect
\refstepcounter{Cscr}\protect\label{au:Costella2002a}%
\renewcommand\theCscr{2002a}\protect\refstepcounter{Cscr}\protect
\label{yr:Costella2002a}{}%
\verb+hep+%
\verb+-lat/0207008+.\addtocounter{CBcit}{1}\item[\hfill{[}\arabic
{CBcit}{]}]\renewcommand\theCscr{\arabic{CBcit}}\protect
\refstepcounter{Cscr}\protect\label{cit:Costella2002b}J.~P.~Cost%
ella, \renewcommand\theCscr{Costella}\protect\refstepcounter{Csc%
r}\protect\label{au:Costella2002b}\renewcommand\theCscr{2002b}%
\protect\refstepcounter{Cscr}\protect\label{yr:Costella2002b}{}%
\verb+hep+%
\verb+-lat/0207015+.\addtocounter{CBcit}{1}\item[\hfill{[}\arabic
{CBcit}{]}]\renewcommand\theCscr{\arabic{CBcit}}\protect
\refstepcounter{Cscr}\protect\label{cit:Drell1976a}S.~D.~Drell, 
M.~Weinstein\ and S.~Yankielowicz, \renewcommand\theCscr{Drell, 
Weinstein\ and Yankielowicz}\protect\refstepcounter{Cscr}\protect
\label{au:Drell1976a}\renewcommand\theCscr{1976a}\protect
\refstepcounter{Cscr}\protect\label{yr:Drell1976a}\mbox{}\protect
\/{\protect\em Phys. Rev.~D\protect\/}\ {\bf14}\ (1976) 487.%
\addtocounter{CBcit}{1}\item[\hfill{[}\arabic{CBcit}{]}]%
\renewcommand\theCscr{\arabic{CBcit}}\protect\refstepcounter{Csc%
r}\protect\label{cit:Drell1976b}S.~D.~Drell, M.~Weinstein\ and S%
.~Yankielowicz, \renewcommand\theCscr{Drell, Weinstein\ and Yank%
ielowicz}\protect\refstepcounter{Cscr}\protect\label{au:Drell197%
6b}\renewcommand\theCscr{1976b}\protect\refstepcounter{Cscr}%
\protect\label{yr:Drell1976b}\mbox{}\protect\/{\protect\em Phys. 
Rev.~D\protect\/}\ {\bf14}\ (1976) 1627.\renewcommand\theCscr{%
\arabic{CBcit}}\protect\refstepcounter{Cscr}\protect\label{citla%
st}\settowidth\Cscr{~[\ref{cit:Drell1976b}]}\end{list}\par\end{d%
ocument}\par
%